\documentclass[aps,prl,twocolumn,showpacs,superscriptaddress]{revtex4}
\usepackage{amsfonts}
\usepackage{amssymb}
\usepackage{amsmath}
\usepackage[dvips]{graphicx}
\usepackage{bm}

\def\diff{\mathrm d}
\def\kB{k_{\mathrm B}}

\def\Q6{Q_{\mathrm 6}}
\def\shearrate{\dot{\gamma}}
\def\shearrateex{\dot{\gamma}^{\mathrm{ex}}}

\newcommand{\bra}{\left\langle}
\newcommand{\ket}{\right\rangle}

\begin{document}
\title{Shear-induced criticality near a liquid-solid transition
of colloidal suspensions}
\date{\today}
\author{Masamichi J. \surname{Miyama}}
\affiliation{%
Department of Pure and Applied Sciences, University of Tokyo, Komaba,
Meguro-ku, Tokyo 153-8902, Japan\\
}%
\author{Shin-ichi \surname{Sasa}}
\affiliation{%
Department of Pure and Applied Sciences, University of Tokyo, Komaba,
Meguro-ku, Tokyo 153-8902, Japan\\
}%

\begin{abstract}
We investigate colloidal suspensions under shear flow through numerical
experiments. By measuring the time-correlation function 
of a bond-orientational order parameter, we find a divergent 
time scale near a transition point from a disordered fluid phase 
to an ordered fluid phase, where the order is characterized by 
a nonzero value of the bond-orientational order parameter. 
We also present a phase diagram in the $(\rho, \shearrateex)$ plane, 
where $\rho$ is the density of the colloidal particles and $\shearrateex$ 
is the shear rate of the solvent. 
The transition line in the phase diagram terminates at the 
equilibrium transition point, while a critical region 
near the transition line vanishes 
continuously as $\shearrateex \rightarrow 0$. 
\end{abstract}
\pacs{64.60.Cn, 83.10.Mj, 83.50.Ax, 83.60.Rs}

\maketitle

\paragraph{Introduction:}


A crystal phase is distinguished from a liquid phase by 
a translational and rotational symmetry breaking in space.
Since there exists an order parameter associated with the 
symmetry-breaking, the nature of the transition to a crystal phase 
is rather different from that of gas-liquid transitions, 
although both are first-order transitions from the viewpoint
of thermodynamics. In short, the transition to a crystal 
phase is classified as a symmetry-breaking first-order transition.


The correlation time of fluctuations 
does not diverge near symmetry-breaking first-order transitions.
This is in sharp contrast to the case of a symmetry-breaking 
second-order transition, which exhibits a divergent time scale,
as is observed in the case of decreasing temperature around a critical point 
of liquid-gas transitions. On the basis of standard understanding,
in the present Letter, we argue that a divergent time
scale of steady-state fluctuations may appear near an equilibrium 
crystallization  (symmetry-breaking first-order transition) point
when a non-equilibrium condition is imposed on the system. 


Concretely, we study colloidal suspensions under shear flow 
through numerical experiments. Since the pioneering work by Ackerson 
and Clark \cite{AckersonClark}, there have been extensive studies 
related to the crystallization of  colloidal suspensions under shear 
flow \cite{ButlerHarrowell, Dhont:2005, ButlerHarrowell:2003,
FrenkelLoewen,MiyamaSasa}. In particular, a phase diagram was 
obtained by numerical experiments \cite{ButlerHarrowell} and laboratory 
experiments \cite{Dhont:2005}, together with numerical realizations 
of  crystal-liquid coexistence under shear flow \cite{ButlerHarrowell:2003}. 
In order to clarify the microscopic mechanism for the transition to
a crystal, the kinetics of homogeneous nucleation under shear flow 
was also investigated \cite{FrenkelLoewen}. However, to our knowledge,
shear-induced criticality near an equilibrium crystallization point has never
been reported, except for our preliminary observation \cite{MiyamaSasa}.


In the present study, we focus on time scales associated 
with the relaxation from a bond-orientational ordered state in a 
disordered regime. We first note that the relaxation  
time, even in equilibrium cases, diverges near the 
melting point. We characterize the parameter dependence 
of the relaxation time quantitatively by measurement of the time 
series of a bond-orientational order parameter. The result is well fitted by
the Vogel-Fulcher law, which suggests the existence of a nucleation 
process of disordered fluid regions in a crystal state. However, since 
crystals do not appear spontaneously in the disordered phase, 
the divergent time scale has never been observed in steady state 
fluctuations. In contrast, in non-equilibrium systems under shear flow, 
the relaxation time exhibits a power-law divergence near a transition point to
an ordered fluid, which suggests
the existence of critical slowing down. The power-law
divergent time scale is also observed in steady state fluctuations.
We refer to this novel phenomenon 
as {\it shear-induced criticality near a liquid-solid transition}.

\paragraph{Model:}           %


We investigate $N$ colloidal particles that are 
suspended in a solvent fluid confined to an 
$L \times L \times L$ cubic box. We impose planar 
Couette flow on the solvent and choose 
the $x$-axis and the $z$-axis to be the directions of 
the shear velocity and the velocity gradient, respectively. 
Concretely, the velocity profile of the solvent is 
assumed to be given as $(\shearrateex z, 0, 0)$. 
We impose periodic boundary conditions along the $x$-axis and 
the $y$-axis and introduce two parallel walls 
so as to confine particles in the $z$ direction. 

Let $\bm{r}_i$, $i=1,\cdots,N$, be the position of particle $i$.
The potential energy of particles $U(\{\bm{r}_j\}_{j = 1}^N) $ consists 
of two parts, $\sum_{i < j} U^{\mathrm{LJ}}(|\bm{r}_i-\bm{r}_j|)$ and
$\sum_{i} U^{\mathrm{wall}}(\bm{r}_i)$. 
The former describes the interaction potential among particles, where
$U^{\mathrm{LJ}}(r) = 4 \epsilon ((\sigma/r)^{12} 
- (\sigma/r)^6) - U_{\mathrm{cutoff}}$
for $r < r_{\mathrm{c}}$ 
with cut-off length $r_{\mathrm{c}}$ 
and 
$U_{\mathrm{cutoff}} = 
4 \epsilon ((\sigma / r_{\mathrm{c}})^{12} - (\sigma / r_{\mathrm{c}})^6)$, 
while 
$U^{\mathrm{LJ}}(r) = 0$ otherwise. 
$U^{\mathrm{wall}}(\bm{r}_i)$ represents the wall potential and is given by 
$U^{\mathrm{wall}}(\bm{r}_i) = u^{\mathrm{WCA}}(r^* - L/2 \pm z_i)$ for $L/2 \pm
z_i < r^*$ with the Weeks-Chandler-Andersen potential 
$u^{\mathrm{WCA}}$ \cite{WCA}, while $U^{\mathrm{wall}}(\bm{r}_i) = 0$ otherwise.


We define momentum of the $i$-th particle relative 
to the shear flow as $\bm{p}_i(t) \equiv m \dot{\bm{r}}_i(t) - m 
\shearrateex z_i(t)\bm{e}_x$, where $m$ is the mass of a 
single particle. We then assume the equation of motion for the particles as
\begin{equation}
 \displaystyle \frac{\diff \bm{p}_i}{\diff t} 
  = \displaystyle -\frac{\partial U(\{\bm{r}_j\}_{j=1}^N)}
  {\partial \bm{r}_i} 
  - \zeta \frac{\bm{p}_i}{m} + \bm{\xi}_i(t), 
\label{eq:genlangevin}
\end{equation}
where $\bm{\xi}_i = (\xi^x_i, \xi^y_i, \xi^z_i)$ represents 
thermal noise satisfying $\langle \xi^{\alpha}_i(t) \xi^{\beta}_j(t') 
\rangle = 2\zeta \kB T \delta_{ij}\delta^{\alpha\beta}
 \delta(t - t')$, $\kB$ is the Boltzmann constant, $T$ is the temperature of the solvent,
and $\zeta$ is the friction coefficient. The superscripts $\alpha$ 
and $\beta$ represent Cartesian components. (See Ref.~\cite{evans:2001}.)


In numerical simulations, all the quantities are converted to 
dimensionless forms by 
setting $m = \sigma = \epsilon = 1$. 
We fix $T = 1.5$, $\zeta / \kB T = 1$, $N=1024$, $r_{\mathrm c} = 2.5 \sigma$,
and $r^* = 0.5 \sigma$, and treat $\rho$ and $\shearrateex$ 
as control parameters. We discretize (\ref{eq:genlangevin}) according to 
the reversible system propagator algorithm method \cite{respa} with time step 
$\Delta t =1 / 256$. In the present Letter, 
$\left< \cdots \right>$ represents the statistical average 
in steady states. 

\paragraph{Order parameters:}    %


We define an order parameter that characterizes a rotational 
symmetry breaking \cite{Q6,crystal}.
Using the Delaunay triangular decomposition \cite{decom} 
on a particle configuration, we determine neighboring particles
for a given particle $i$. The collection of edges that extend from 
$\bm{r}_i$ in the Delaunay triangular is denoted by 
$(\bm{r}_{ij})_{j=1}^{n_{\mathrm{B}}(i)}$, where $n_{\mathrm{B}}(i)$
represents the number of neighbors of particle $i$.
From this collection, we define a 13-dimensional vector 
$\bm{q}_6(i) = (q_{6,-6}(i), \dots,q_{6,m}(i), \dots q_{6,6}(i))$
as 
\begin{equation}
q_{6,m}(i) = \frac{1}{n_{\mathrm{B}}(i)}
\sum_{j=1}^{n_{\mathrm{B}}(i)} Y_{6,m}
\left( \frac{\bm{r}_{ij}}{|\bm{r}_{ij}|}  \right),
\end{equation}
where $Y_{\mathrm{6m}}$ is the spherical harmonics function of 
degree six. Then, the bond-orientational order is 
qualified by 
\begin{equation}
\bar q_{6,m}=\frac{1}{N}\sum_{i=1}^N q_{6,m}(i).
\end{equation}
Here, $\bra \bar q_{6,m}\ket=0 $ if the rotational symmetry 
is not broken, while $\bra \bar q_{6,m}\ket \not =0 $ in the 
thermodynamic limit $N \to \infty$ when the bond-orientational
order emerges.
In order to detect symmetry-breaking, it is convenient 
to measure  the magnitude of the vector $\bar q_{6,m}$. Following
the standard convention, we define
\begin{equation}
Q_6 \equiv \bra \left( \frac{4\pi}{13} \sum_{m=-6}^6 
\bar q_{6,m}\bar q_{6,m}^*  \right)^{1/2} \ket. 
\end{equation}
Note that $Q_6 \simeq O(1/\sqrt{N})$ in the disordered phase,
while $Q_6 \simeq O(1)$ in the ordered phase, when $N \to \infty.$

\begin{figure}[tbp]
\includegraphics[width=8truecm,clip]{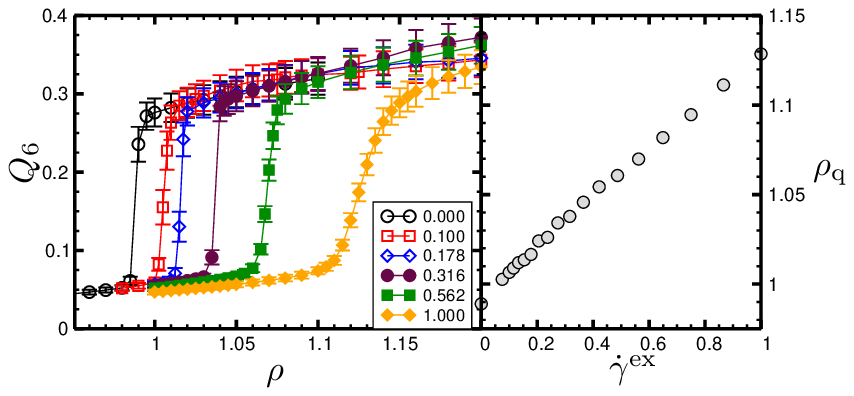} 
\caption{
(Color online) Left: $Q_6$ as a function of $\rho$ for several
values of $\shearrateex$.
Right: $\rho_{q}$ as a function of $\shearrateex$.} 
\label{fig1}
\end{figure}

On the left-hand side of Fig.~\ref{fig1}, we show $Q_6$ as a function 
of $\rho$ for several values of $\shearrateex$. The figure indicates the existence of an ordered state with $Q_6 \simeq O(1)$ in a high-density regime for each $\shearrateex$. In particular, 
the transition to the ordered phase is quite sharp when $\shearrateex=0$, 
while the transition width becomes wider as $\shearrateex$ 
is increased. For a tentative value of the transition point, 
we define $\rho_{q}$ as the density such that $Q_6 =0.2$. 
We display $\rho_{q}$ as a function of $\shearrateex$ 
in the right-hand side of Fig.~\ref{fig1}.  More precise determination
of the functional forms of $Q_6$ will be obtained by investigating 
larger systems.  Note that in the thermodynamic limit without shear flow, 
non-zero $Q_6$ emerges continuously for a density at which a crystal 
can coexist with a liquid. When we ignore the coexistence phase,
$Q_6$ exhibits a discontinuous transition, which is 
observed for the system under constant pressure. In the present Letter, 
putting aside phenomena associated with the coexistence phase,
we focus on the question of how the nature of the symmetry-breaking 
discontinuous-transition is modified by the influence of shear flow.

\begin{figure}
\includegraphics[width=7.5truecm,clip,bb=140 329 342 452]{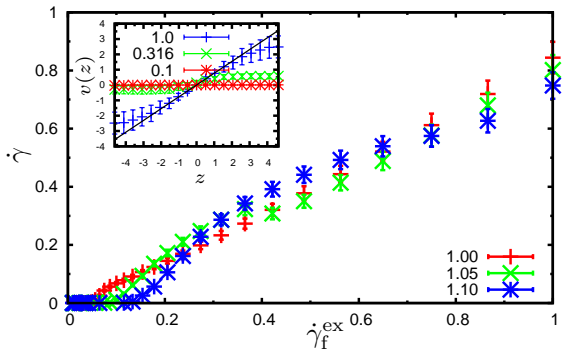}
\caption{(Color online) Shear rate of particle flow as a function of 
$\shearrateex$ for several values of $\rho$. 
Each point is obtained from the velocity profiles $\bar{v}(z)$. 
Examples of velocity profiles for different $\shearrateex$ 
with $\rho = 1.1$ fixed are shown in the inset.
}
\label{fig2}
\end{figure}

\begin{figure}
\begin{center}
\includegraphics[width=7.5truecm,clip,bb=140 328 346 455]{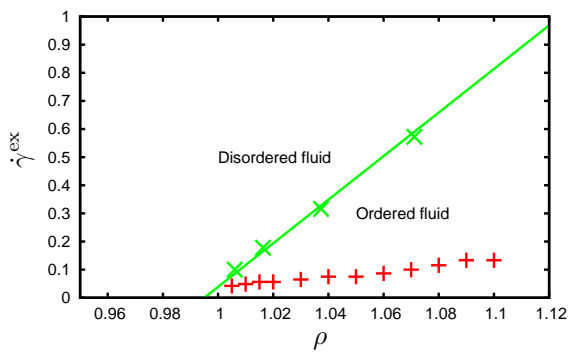}
\caption{(Color online) 
Phase diagram in the $(\rho,\shearrateex)$ plane.
The green and cross symbols represent the transition line 
$\rho=\rho_q(\shearrateex)$ between the disordered fluid phase 
and the ordered fluid phase.  
The parameter values for realizing $\shearrate=10^{-3}$, 
below which flow appears to cease, 
are also plotted as red plus symbols.}
\label{fig3}
\end{center} 
\end{figure}

In the equilibrium case, the ordered phase corresponds to 
a crystal. However, in the non-equilibrium cases, since the shear 
flow drives particles, particles may flow even in the ordered phase 
with $Q_6 \not = 0$. We then measure the $x$-component of 
the velocity averaged over a region with an interval $[z+0.5, z-0.5]$ 
in the $z$ direction, which is denoted by $\bar{v}(z)$.  
Examples of $\bar{v}(z)$ for several values of $\shearrateex$ 
with $\rho=1.1$ fixed are shown in the inset of 
Fig.~\ref{fig2}. We then determine the shear rate 
$\shearrate$ of particles by fitting the slope of 
the velocity profile $\bar{v}(z)$ in the region 
$-1.5 < z < 1.5$. The obtained shear rates $\shearrate$ are
plotted for $\shearrateex$ in Fig.~\ref{fig2}. Although the
flow might cease at some value of 
$\shearrateex$, the determination as to whether the cross-over 
is actually singular is a delicate problem.
 (See Ref. \cite{Biroli} for a related discussion.)
For any case, the cross-over points are located at a
higher density than $\rho_q$ when $\shearrateex >0$.
(See Fig.~\ref{fig3}.) 
Therefore, as we are concerned with behaviors
near the transition point at which the order parameter 
$Q_6 \simeq O(1)$ appears, we may assume that 
an ordered fluid is observed in the ordered phase 
in the non-equilibrium cases. 

\paragraph{Transition to the ordered fluid:}    %

\begin{figure}[tbp]
\includegraphics[width=7.5truecm,clip,bb=148 330 346 453]{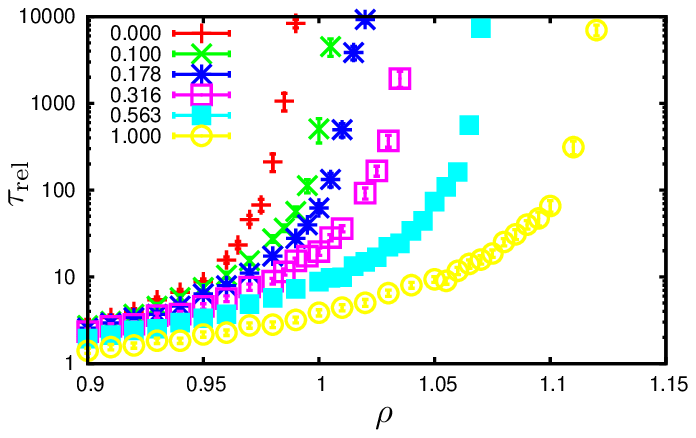} 
\caption{(Color online) 
Relaxation time $\tau_{\mathrm{rel}}$ as a function of 
density $\rho$ for several values of $\shearrateex$.} 
\label{fig4}
\end{figure}

Next, we characterize the nature of the  
transition to the ordered phase. First, in order to observe 
a clear difference between the equilibrium and non-equilibrium cases,
we measure the relaxation time $\tau_{\mathrm{rel}}$ at which $Q_6$ 
reaches a value of 0.05, starting from a crystal state, 
which $Q_6 \approx 0.35$. Note that $\tau_{\mathrm{rel}}$ 
can be measured only in the disordered phase.  In Fig.~\ref{fig4}, we show $\tau_{\mathrm{rel}}$ as a function of $\rho$ for several values of $\shearrateex$, where we set the maximum waiting time to $\tau = 10000$. The results indicate the existence of a characteristic density $\rho_{\rm d}$ at which the relaxation time diverges for each value of $\shearrateex$. 

\begin{figure}[tbp]
\includegraphics[width=7.5truecm,clip,bb=143 332 341 450]{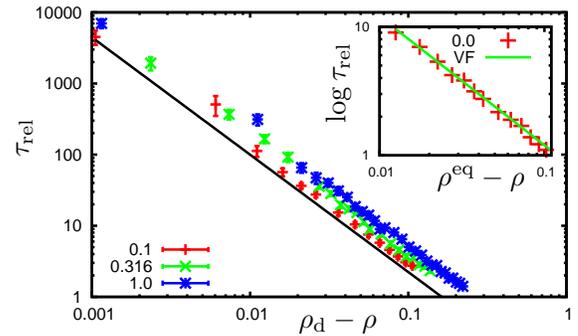} 
\caption{(Color online) 
Fitting of functional forms of $\tau_{\mathrm{rel}}$. 
$\tau_{\mathrm{rel}}$ versus 
$\rho_{\rm d}-\rho$  with a log-log plot for 
$\shearrate^{\rm ex}=0.1$, $0.316$, and $1.0$. 
$\rho_{\rm d}$ is a fitting parameter, the value of which is estimated
as 1.006, 1.037, and 1.121, respectively.
The guide line represents (\ref{power-law}) with 
$\zeta=1.6$. Inset: $\log\tau_{\mathrm{rel}}$ versus $\rho_{\rm d}-\rho$ with a
 log-log plot for $\shearrate^{\rm ex}=0$. The guide line represents
(\ref{VF}) with $A=0.1$ and $\rho_{\rm d}=1.003$. 
}
\label{fig5}
\end{figure}
\begin{figure}[tbp]
\includegraphics[width=7truecm,clip,bb=134 326 341 454]{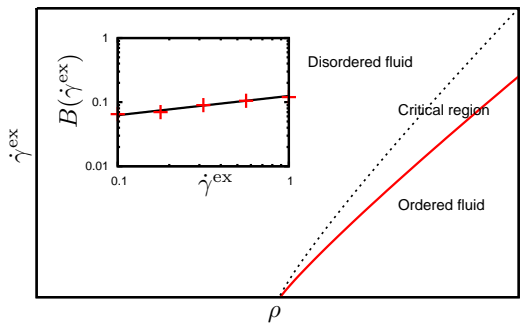} 
\caption{(Color online) Schematic phase diagram with the critical region 
in the ($\rho$-$\shearrateex$) plane. Inset: $B(\shearrateex)$ 
as a function of $\shearrateex$. The guide line represents 
 a power-law function with exponent 0.3. 
}
\label{fig6}
\end{figure}

Let us determine the functional form of $\tau_{\mathrm{rel}}$
with the value of $\rho_{\mathrm{d}}$.
First, as shown in the inset of Fig.~\ref{fig5}, 
$\tau_{\mathrm{rel}}$ for the equilibrium case
is well fitted by the Vogel-Fulcher law
\begin{equation}
\tau_{\mathrm{rel}}
\simeq \tau_0\exp\left(\frac{A}{\rho_{\mathrm{d}}-\rho} \right).
\label{VF}
\end{equation}
A phenomenological argument may be developed for the 
nucleation of a disordered domain, by which (\ref{VF})
may be understood. (See for example Ref.~\cite{Krzakala} 
for a demonstration of a $q$-states Potts model.)
In contrast, Fig.~\ref{fig5} indicates that
$\tau_{\mathrm{rel}}$ for the systems
under shear flow follows a power-law form
\begin{equation}
\tau_{\mathrm{rel}}
\simeq B(\shearrateex)
(\rho_{\mathrm{d}}(\shearrateex)
-\rho)^{-\zeta},
\label{power-law}
\end{equation}
where $\zeta \simeq 1.6$. 
This suggests that the divergent behavior does not originate from 
the nucleation of disordered regions but may
be related to critical slowing down. 
Note that the pre-factor $B$ in (\ref{power-law})
depends slightly on $\shearrateex$ in the form 
$B\simeq (\shearrateex)^{0.3}$, as shown in 
the inset of Fig.~\ref{fig6}. Let $\rho_{\rm w}
(\shearrateex)$ be a typical width of 
the power-law region for $\shearrateex $. Then, on the basis of 
the dimensional analysis, it is expected that 
\begin{equation}
\shearrateex
\tau_{\mathrm{rel}}
\simeq 
\left(
\frac{\rho_{\mathrm{d}}(\shearrateex)-\rho}
{\rho_{\rm w}}
\right)
^{-\zeta}.
\label{power-law2}
\end{equation}
By assuming $\rho_{\rm w} \simeq (\shearrateex)^{\chi}$
in (\ref{power-law2}),
we obtain $B \simeq (\shearrateex)^{\chi\zeta-1}$, which leads to
$\chi\simeq 1.3/1.6 > 0$. This means that 
a critical region for the system with finite $\shearrateex$
becomes narrower for smaller $\shearrateex$
and vanishes in the equilibrium system. See the schematic 
phase diagram in Fig.~\ref{fig6}.

\begin{figure}[tbp]
\includegraphics[width=4.25truecm,clip,bb=159 331 285 451]{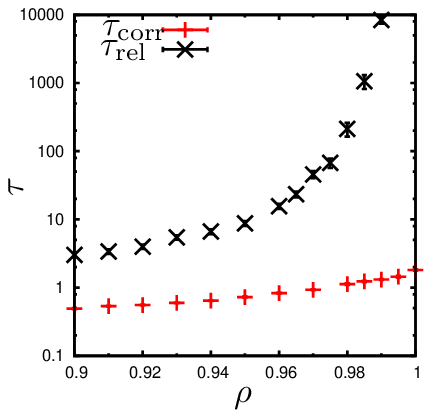} 
\includegraphics[width=4.25truecm,clip,bb=159 331 285 451]{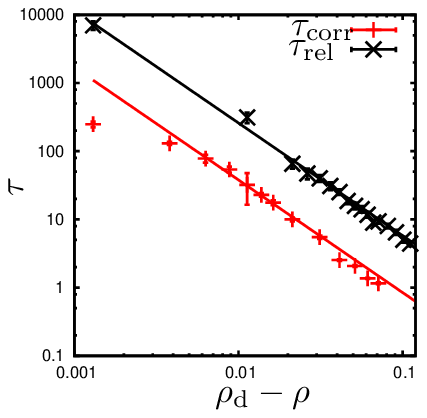} 
\caption{(Color online) 
Correlation time $\tau_{\mathrm{corr}}$ in steady states 
as a function of $\rho$. The melting time $\tau_{\mathrm{rel}}$  is also 
superimposed for comparison. 
(Left:) $\shearrateex = 0$, 
and
(Right:) $\shearrateex = 1.0$.
}
\label{fig7}
\end{figure}

We now note that such a divergent time scale
is never observed in the stationary state of the 
equilibrium system. In order to confirm this 
explicitly, we measure the time correlation function defined by 
\begin{equation}
C(t)= \sum_{m=-6}^6 
[\bra \bar q_{6,m}(t_0) \bar q_{6,m}^*(t_0+t) \ket
-\bra \bar q_{6,m} \ket^2].
\end{equation}
We determine the correlation time $\tau_{\rm c}$  from the fitting of the exponential decay rate 
of $C(t)$. On the left-hand side of Fig.~\ref{fig7}, $\tau_{\rm c}$ is shown as a function of $\rho$ in the disordered regime. Indeed, the correlation time does not diverge. 
For reference, we superimpose the data of the relaxation time $\tau_{\mathrm{rel}}$. 
In contrast to the equilibrium case, as shown in the right-hand side 
of Fig.~\ref{fig7}, the correlation time in the system 
under shear flow diverges in a manner similar to $\tau_{\mathrm{rel}}$. These results indicate that the symmetry-breaking transition to the ordered fluid
accompanies a critical phenomenon. This is the main claim of the present Letter.


\paragraph{Concluding remarks:}       %


Before ending this Letter, we address three considerations.
First, we conjecture that the fluctuation of $\bm{q}_6$
possesses a divergent length scale.
By investigating the manner of divergences of 
several quantities for systems of different sizes,
the universality class for this phenomenon may be determined.


Second, with regard to the universality problem, 
we are also interested in a simple mathematical model in the 
same universality class. For example, it might be 
possible to propose a model describing a stochastic time evolution of  
the coarse-grained order parameter field. The first problem 
in this direction is to derive the value of $\zeta$ using a phenomenological argument.


Finally, in all of the arguments presented above,
the coexistence phase is ignored. In order to
extract more precise results, it might be better to
investigate systems under constant pressure.
A study of such systems of larger 
sizes will be performed in the future.


In summary, we have investigated colloidal suspensions
under shear flow. We have found that the critical transition
line starts from the liquid-solid transition point in 
the equilibrium system without shear flow.
This novel phenomenon, referred to as shear-induced criticality,
will be investigated from several viewpoints.

We thank M. Kobayashi for discussions on the role of 
the order parameters $\bar{q}_{6,m}$. The present study was supported 
by grants from the Ministry of Education, 
Culture, Sports, Science, and Technology 
of Japan, Nos. 21015005 and 22340109,
and by a Grant-in-Aid for JSPS Fellows (DC2), 21-10700, 2009.


\end{document}